\documentclass[12pt]{article}
\usepackage{amssymb, amsmath}
\usepackage{ulem}
\usepackage{amsfonts}
\usepackage{graphicx}
\usepackage{setspace} 
\usepackage{rotating}
\usepackage[utf8]{inputenc}
\usepackage[sectionbib]{natbib}
\usepackage{array,epsfig,fancyheadings,rotating}
\usepackage{caption}
\usepackage{subcaption}
\usepackage{amsmath}
\usepackage{hyperref}
\hypersetup{
  colorlinks=true,
  linkcolor=blue,
  urlcolor=blue,
  citecolor=blue,
}
\usepackage{amsmath}
\usepackage{amssymb}
\usepackage{amsfonts}
\usepackage{multirow}
\usepackage{xcolor}
\usepackage{amsthm}
\usepackage{lscape}
\usepackage{afterpage}
\usepackage{enumitem}
\usepackage{booktabs}
\usepackage{algorithm}
\usepackage{arydshln}
\usepackage{algpseudocode}
\usepackage{authblk}
\theoremstyle{definition}

\newtheorem{remark}{Remark}

\newcommand{\bB}{\mathbf{B}}

\newcommand{\bW}{\mathbf{W}}

\newcommand{\bu}{\mathbf{u}}

\newcommand{\by}{\mathbf{y}}

\newcommand{\sD}{\mathcal{D}}

\newcommand{\sF}{\mathcal{F}}

\newcommand{\sL}{\mathcal{L}}

\newcommand{\sP}{\mathcal{P}}

\newcommand{\sR}{\mathcal{R}}

\newcommand{\sU}{\mathcal{U}}

\newcommand{\bbP}{\mathbb{P}}

\newcommand{\bbR}{\mathbb{R}}

\newcommand{\sfB}{\mathsf{B}}

\newcommand{\sfW}{\mathsf{W}}

\newcommand{\E}{\mathbb{E}}

\newcommand{\tp}{\text{T}}

\DeclareMathOperator{\vect}{vec}

\newcommand{\btheta}{\boldsymbol{\theta}}

\newcommand\numberthis{\addtocounter{equation}{1}\tag{\theequation}}

\title{Scalar-on-distribution regression via generalized odds with applications to accelerometry-assessed disability in multiple sclerosis}

\author[1,$\ast$]{Pratim Guha Niyogi}
\author[2]{Muraleetharan Sanjayan}
\author[2,3]{Kathryn C. Fitzgerald}
\author[2,3]{Ellen M. Mowry}
\author[4]{Vadim Zipunnikov}

\affil[1]{Department of Data Science, University of Mississippi Medical Center}

\affil[2]{Department of Neurology, Johns Hopkins University}

\affil[3]{Department of Epidemiology, Johns Hopkins University}

\affil[4]{Department of Biostatistics, Johns Hopkins University}

\affil[*]{Corresponding author: pguhaniyogi@umc.edu}

\begin{document}
\maketitle

\begin{abstract}
Distributional representations of data collected using digital health technologies have been shown to outperform scalar summaries for clinical prediction, with carefully quantified tail-behavior often driving the gains. Motivated by these findings, we propose a unified generalized odds (GO) framework that represents subject-specific distributions through ratios of probabilities over arbitrary regions of the sample space, subsuming hazard, survival, and residual life representations as special cases. We develop a scale-on-odds regression model using spline-based functional representations with penalization for efficient estimation. Applied to wrist-worn accelerometry data from the HEAL-MS study, generalized odds models yield improved prediction of Expanded Disability Status Scale (EDSS) scores compared to classical scalar and survival-based approaches, demonstrating the value of odds-based distributional covariates for modeling DHT data.
\end{abstract}

\section{Introduction}
Digital health technologies (DHT) collect continuous data on individual behavior, producing rich distributions of measurements such as minute-level activity counts. These distributions contain substantially more information than common scalar summaries like total daily activity count (TAC). A central open question is which distributional representations are most informative, particularly when rare or extreme values carry an important clinical signal.
\par
Recent work by \citet{niyogi2024hazard} provides strong motivation for focusing on tails of the subject-specific distributions. By comparing several subject-level representations of accelerometry data including densities, quantiles, survival, and hazard functions they showed that hazard-based representations achieved markedly better prediction of disability in multiple sclerosis. This advantage was attributed to the hazard function’s emphasis on extreme activity levels rather than typical activity levels. These results provide a key motivation: tail behavior, reflecting high-intensity activity, can be a critical driver of clinical outcomes. 

Building on this motivation, we introduce subject-specific generalized odds functions as distributional covariates. Intuitively, generalized odds functions compare the probability of extremely high activity levels to the probability of more typical activity levels, directly quantifying tail behavior. We demonstrate that they generalize survival and hazard-based distributional objects while remaining well defined for sparse, bounded, or discrete data. Building on this idea, we propose a scalar-on-GO regression framework that uses GO functions as predictors of clinical outcomes. We develop the corresponding methodology and demonstrate, using wearable physical activity data from individuals with multiple sclerosis, that generalized odds substantially improve predictive performance and provide additional clinically actionable information.

\section{Data description}
\label{sec:study}

\subsection{Participants}
We demonstrate a baseline study at the Johns Hopkins University \citep{rjeily2022using, tian2020longitudinal, niyogi2024hazard, fitzgerald2025within} that was conducted from January 2021 till March 2023 on a cohort of longitudinal observation set-up conducted by Home-based Evaluation of Actigraphy to Predict Longitudinal Function in Multiple Sclerosis (HEAL-MS). Multiple sclerosis (MS) is a complex and challenging neurological disorder that affects the central nervous system, primarily the brain and the spinal cord. It is characterized by a wide range of symptoms and can have a profound impact on a person's quality of life. Due to its unpredictable nature and the diversity of symptoms it presents, MS has been the subject of extensive research and study. MS research seeks to unravel the intricate mechanisms underlying the disease, discover more effective treatments, and ultimately find a cure. The disease's etiology remains a subject of investigation, with a consensus that it involves both genetic and environmental factors. The immune system's misguided attack on the protective myelin sheath surrounding nerve fibers is a key pathological feature. The expanded disability status scale (EDSS) is a standard clinical measure to detect MS. This EDSS score is non-linear and provides a total score on a scale that ranges from 0 to 10. Normal neurological examination ends up with EDSS score of zero. In this study, the main goal is to establish the relationship between EDSS scores and changes in physical activity. 
The subjects of the sample are older (age greater than 40 years, the mean age is 54.8 years with a standard deviation 8.5; 70.5\% are female) without known comorbidities that may limit physical activity. A total of 248 samples were included in the analyses. 

\subsection{Physical activity}
An accelerometer, a safe and inexpensive tool was used to measure the disability in people with MS. Accelerometry metrics were captured using GT9X Actigraph which measures the activity using triaxial technology to obtain activity counts in 1-minute epochs. All participants were asked to wear the actigraph device on the wrist of their non-dominant hand for the entire day over a duration of 2 weeks. Accelerometers were set to capture three-dimensional acceleration at a 30Hz sampling rate and raw data (.gt3x) were downloaded from the device using ActiLife v6.134 Lite edition. Binary raw activity data were read by \texttt{read.gt3x} package in R that creates a data frame of physical activity (PA) in a 1440 analytic format. To assure the quality of the physical activity measures, we further proceed the following steps: (i) A non-wear minute is categorized when it is associated with any continuous 90-minute intervals of zero values in minute-level activity counts data, (ii) We declare the valid days if wear-times is greater than 90\% of the day, (ii) each participant have least 3 valid days of activity data. For the analysis of the data using the proposed methodology, we restrict our attention to such 1440 subject-specific minute-level physical activity counts from 8am to 8pm so that we can ignore a general non-active time for each subject on an average. In addition, the activity counts are log-transformed using the transformation $x \rightarrow \log (x+1)$ to remove the potential skewness of the data.

\section{Model}
\label{sec:model}
\subsection{Statistical framework}
\label{sec:framework}
Assume that $\by = (y_{1}, \cdots, y_{n})^{\tp} \in \bbR^{n}$ denotes the collection of random variables from an exponential family of size $n$ and we observe the collection of scalars $X_{i1}, \cdots, X_{im_{i}}$ for $i = 1, \cdots, n$ where $\{ X_{ij}: j = 1, \cdots, m_{i}\} \sim p_{i}$ and $p_{i}$ is the subject-specific mass function. Throughout this paper, we assume that $X_{ij}$ are non-negative and this assumption is very trivial in our data example. 
In addition to the non-negativity, we assume that the $X$ takes values from the set of a finite number of atoms, such sets are defined as $\sD = \{ 0, 1, \cdots, D\}$. Thus for $u \in \sD$, the survival function is $S(u) = \bbP\{ X > u\} = \bbP\{ X \geq u+1\} = \sum\limits_{k=u+1}^{D}p(k)$ where $p(k)$ is the probability mass function of $X$ at point $k$ and the distribution of $X$ is defined as $F(u) = \bbP\{X \leq u\} = \sum\limits_{k=1}^{u}p(k)$ such that $F(0) = 0$ and $F(D) = 1$. Furthermore, the hazard function at $u$ is $\lambda(u) = p(u)/S(u) = p(u)/\sum\limits_{k=u+1}^{D}p(k)$ which is atmost 1. Thus, $\sum\limits_{k=1}^{D}\lambda(k) \leq D$ is always finite, and therefore, the hazard function is integrable. Although, $\lambda(u)$ is a function on a discrete domain, it is always possible to assume the smoothness of $\lambda(u)$ over $u$. 
\par
Let us start with the simple scalar-on-distribution regression problem when the underlying distribution object of interests is a subject-specific survival function, viz., $S_{i}(u) = \bbP\{ X_{i} > u\}$. Therefore, based on the paired data $\{y_{i}, S_{i}(\cdot)\}_{i=1}^{n}$, one can use the regression model, 
\begin{equation}
\label{eq:surv}
    y_{i} \sim EF(\mu_{i}, \sigma^{2}); \qquad g(\mu_{i}) = \alpha + \int_{0}^{D} \beta_{S}(u)S_{i}(u)du, 
\end{equation}
where $\mu_{i} = \E\{y_{i}|X_{i1}, \cdots, X_{im_{i}}\}$ and $g$ is a known link function (for example, logit or linear function). In general, the survival function provides insight into the behavior of the right tail of the distribution. In the context of studies on physical activity, the survival function of the activity counts metric quantifies how often an individual's activity level surpasses a specific intensity. If $S(u)$ is large for large $u$, then the high activity is common. An individuals with sedentary behavior, the survival function drops sharply at low values of $u$; whereas the survival function remains higher for larger values of $u$ for an active individual with a higher activity level. In the regression model in Equation \eqref{eq:surv}, $\beta_{S}(u)$ defines the weight given to the survival function at each $u \in [0, D]$ and the inner product between subject-specific survival function $S_{i}$ and $\beta_{S}$ is overall contribution of the survival function in the regression model. At each $u$, $\beta_{S}(u)$ is the effect of outcome for a one-unit increase in $S(u)$ when all other values $S_{i}(u')$ are fixed (for $u' \neq u$). Again, in the context of the physical activity profile, large positive values of $\beta(s)$ for higher values of $u$ indicate a stronger effect of moderate-to-vigorous activity in predicting the response (such as EDSS scores or any other clinical onsets).
\par
If $\beta_{S}$ is a constant function (i.e., $\beta_{S}(u) = \beta_{0}$), then $g(\mu_{i}) = \alpha + \beta_{0}\E\{X_{i}\}$, which is the generalized linear model in classical set-up. On the other hand, one can write further that 
\begin{align*}
    g(\mu_{i}) &= \alpha + \int_{0}^{D}\beta_{S}(u)S_{i}(u)du\\
    &= \alpha + \int_{0}^{D}\beta_{S}(u)\frac{S_{i}(u)}{F_{i}(D)}du \qquad \text{since } F(D) = 1\\
    &= \alpha + \int_{0}^{D}\beta_{S}(u)h_{2, F_{i}}(D, u)du,\numberthis
\end{align*}
where $h_{2, F_{i}}(u_{1}, u_{2}) = \frac{S_{i}(u_{2})}{F_{i}(u_{1})} = \frac{\bbP\{X > u_{2}\}}{\bbP\{X \leq u_{1}\}}$ for $i = 1, \cdots, n$. 
\par
This motivates us to use the ratio of subject-specific probabilities of the two different events to model $y_{i}$. For example, for subject $i$, one can easily generalize the function $h_{2, F_{i}}$ as a ratio of two probabilities $\bbP\{X_{i} \in A\}/\bbP\{X_{i}\in B\}$ for two sets $A, B \subset [0, D]$. This ratio of the probabilities of the two events $\{X_{i} \in A\}$ and $\{ X_{i} \in B\}$ is defined as odds of event $\{X_{i} \in A\}$ against that of $\{X_{i} \in B\}$. The subject-specific quantities $h_{2, F_{i}}$ as a ratio can help in understanding the relative likelihood of observing extreme activity levels (either high or low) compared to more moderate activity levels. Depending on the values of $u_{1}$ and $u_{2}$, this could be used to identify patterns in activity data, such as how likely it is for a subject to experience very high versus very low levels of activity. If $u_{1}$ and $u_{2}$ are the thresholds for the moderate-to-vigorous and sedentary activities respectively, then the ratio $h_{2, F_{i}}$ greater than 1 suggests that high activity levels are more likely than low activity levels. Conversely, if the ratio is less than 1, it suggests that the person is more likely to be inactive or engage in lower activity levels than to experience high-intensity activity.
\par
In addition, when $u_{1} = u_{2} (= u)$, then $h_{2, F_{i}}$ reduces to classical odds ratio for subject $i$ at level $u$. For example, if the two events are ``survival beyond time $u$" and failure by the $u$, then the odds of survival play an important role in the statistical literature with applicability in medical sciences. For clarity, here the word ``survival at certain thresholds" means that the activity counts are above that threshold.
\par
Based on the different choices of $u$s, we define different distributional objects based on the odds that characterize different properties of the underlying distributions in the following subsections. In addition, we use such odds to predict the clinical outcomes using the tools available in functional data analysis and modern machine learning literature. 


\subsection{Modeling with 1-index object}
\label{sec:1index}
For the random variable $X$ discussed in the earlier paragraph, we define the ``odds of survival" beyond the point $u$ as a smooth function of $u$. Let us consider a function $h_{1,F}: [0, D] \rightarrow \bbR_{\geq 0}$ such that 
\begin{equation}
    h_{1, F}(u) = \frac{S(u)}{F(u)} = \frac{1-F(u)}{F(u)}.
\end{equation}
Here, we assume that $h_{1, F_{1}}, \cdots, h_{1, F_{n}}$ are independent and identically distributed (i.i.d.) random functions with the same distribution as a generic $f_{1, F} \in \sF([0, D])$ and this statistical process has continuous covariance function in $[0, D]$, for $D > 0$. To see the effect of such function $h_{1, F_{i}}$ on the clinical outcome $y_{i}$, we consider the linear regression involving the scalar response $y$ and functional predictor $h_{1, F_{i}}$ defined on $[0, D]$. Thus, we assume that the data $(y_{i}, h_{1, F_{i}}(\cdot))_{i=1}^{n}$ are random sample from $(y, h_{1, F})$. 
Further, we assume that $\beta$ is the coefficient function in $\sF([0, D])$. Therefore, we notice that the linear functional $\int_{0}^{D}\beta_{1}(u)h_{1, F_{i}}(u)du$ describes common effects of the entire trajectories $h_{1, F_{i}}(\cdot)$ on $y_{i}$. Therefore, the functional linear model can be stated as 
\begin{equation}
\label{eq:model-1d}
    y_{i} \sim EF(\mu_{i}, \sigma^{2}); \qquad g_{1}(\mu_{i}) = \alpha_{1} + \int_{0}^{D}\beta_{1}(u)h_{1, F_{i}}(u)du
\end{equation}
where $EF(\mu_{i}, \sigma^{2})$ denotes an exponential family distribution with mean $\mu_{i}$ and finite dispersion parameter $\sigma^{2} > 0$. Here $g_{1}$ is the link function and $\alpha_{1}$ is the intercept on the regression model. 
\begin{figure}[h]
    \centering
    \includegraphics[width=0.5\linewidth]{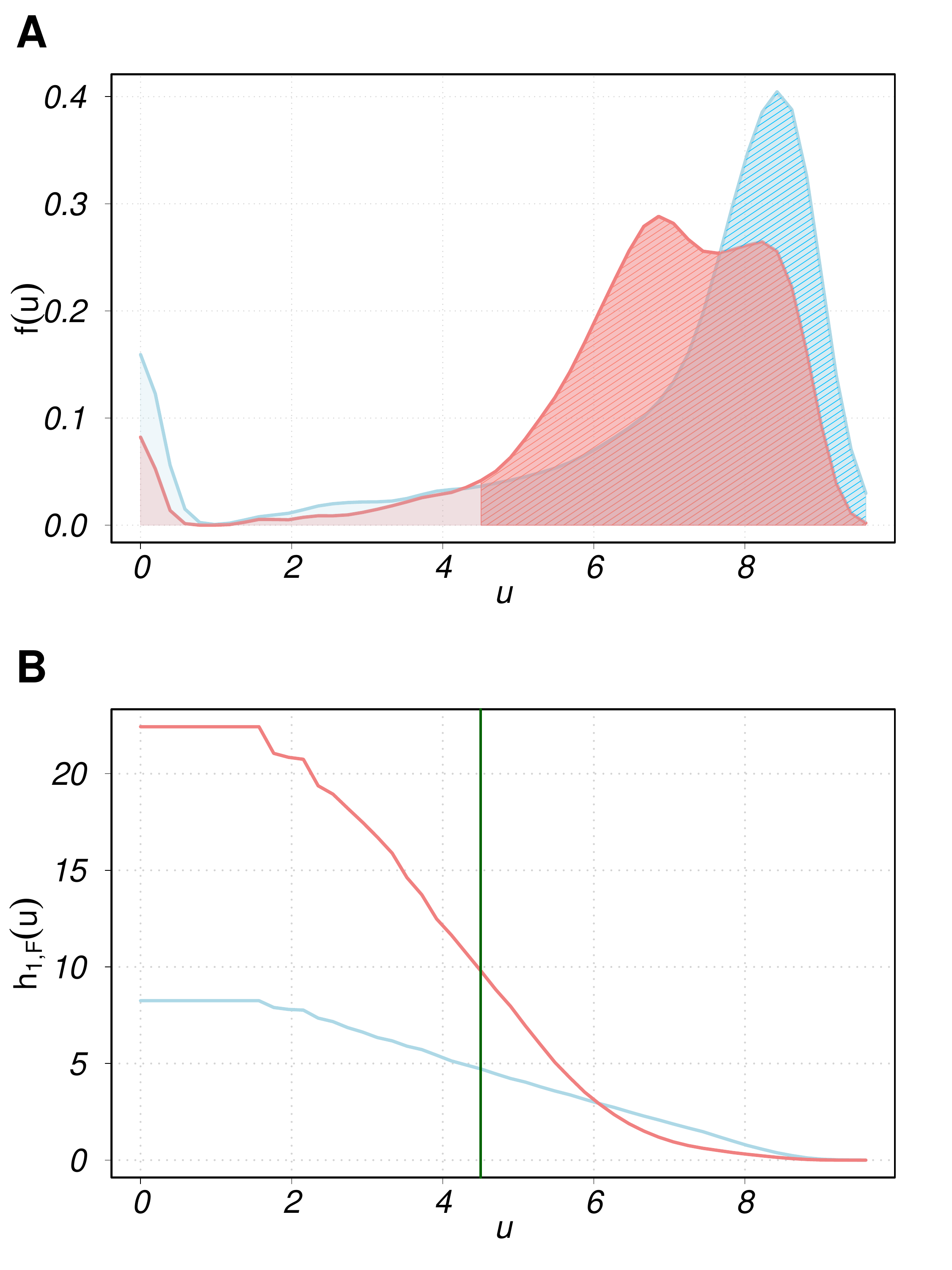}
    \caption{Two randomly selected subjects are selected. Subject 1: 47 years old female with EDSS = 1 (light blue). Subject 2: 62 years old female with EDSS  = 6 (light red). The average log-activity counts for the two subjects are 0.0979 and 0.1007, respectively. Panel A represents the density function of the log-activity counts for two subjects. Panel B represents the 1-index odds $h_{1, F_{i}}$ for subject $i = 1, 2$. For example, we fix $u = 4.5$. The density function below $u$ is similar, but it's much different than the tail of the distribution. $F_{1}(4.5) = 0.17$ and $F_{2}(4.5) = 0.09$, whereas the difference between odds is different ($h_{1, F_{1}}(4.5) = 4.72$ and $h_{1, F_{2}}(4.5) = 9.8$). 
    }
    \label{fig:odds1d}
\end{figure}

In Figure \ref{fig:odds1d}, we represent the odds $h_{1, F_{i}}$ over $u \in [0, D]$ for randomly selected two subjects where the Subject 1 is a 47 years old female with EDSS = 1 and Subject 2 is a 62 years old female with EDSS  = 6. The quantity $h_{1, F_{i}}$ help us to understand the relative likelihood of observing a given activity level; which could be sedentary or moderate to vigorous activity levels. Depending on the values of $u$, this could be used to identify patterns in activity data, such as how likely it is for a subject to experience very high versus very low levels of activity. If $h_{1, F_{i}} > 1$, the probability of being in a high-activity state is greater than the probability of being in a low-activity state. Therefore, the person is more frequently in an active condition than a sedentary or low-activity condition. On the other hand, if $h_{1, F_{i}} < 1$, the probability of being in a low-activity state is greater than the probability of being in a high-activity state. This implies that the person spends more time in lower activity levels than in higher activity levels. However, the intersecting points $\{u: h_{1, F_{1}}(u) = h_{1, F_{2}}(u)\}$ for subject $i=1,2$ could serve as a potential biomarker for clinical studies, which we leave for future research.
\par
Since we assume that the regression coefficient varies smoothly over $u$, we can apply the spline technique that produces a good approximation of the unknown functional coefficient $\beta_{1}$ in terms of the univariate basis function expansion. Thus, we can write a model as follows.
\begin{align*}
\label{eq:g1s}
    g_{1}(\mu_{i}) &= \alpha_{1} + \int_{0}^{D}\beta_{1}(u)
     h_{1, F_{i}}(u)du\\
     &=\alpha_{1} + \int_{0}^{D}\sum_{k=1}^{\kappa}b_{k}^{(1)}B_{k}(u)h_{1, F_{i}}(u)du\\
     &=\alpha_{1} + \sum_{k=1}^{\kappa}b_{k}^{(1)}W_{ik}^{(1)}, \numberthis
\end{align*}
where $W_{ik}^{(1)} = \int_{0}^{D}B_{k}(u)h_{1, F_{i}}(u)du$ for $k = 1, \cdots, \kappa$, and $i = 1, \cdots, n$; $\{B_{k}(u)\}_{k = 1}^{\kappa}$ is the set of known basis functions over $u$. Therefore, the objective is to estimate $b_{k}^{(1)}$s and hence $\beta_{1}$.

\subsection{Modeling with 2-index object}
\label{sec:2index}
\begin{figure}[h]
     \centering
     \begin{subfigure}[b]{0.49\textwidth}
         \centering
         \includegraphics[width=\linewidth]{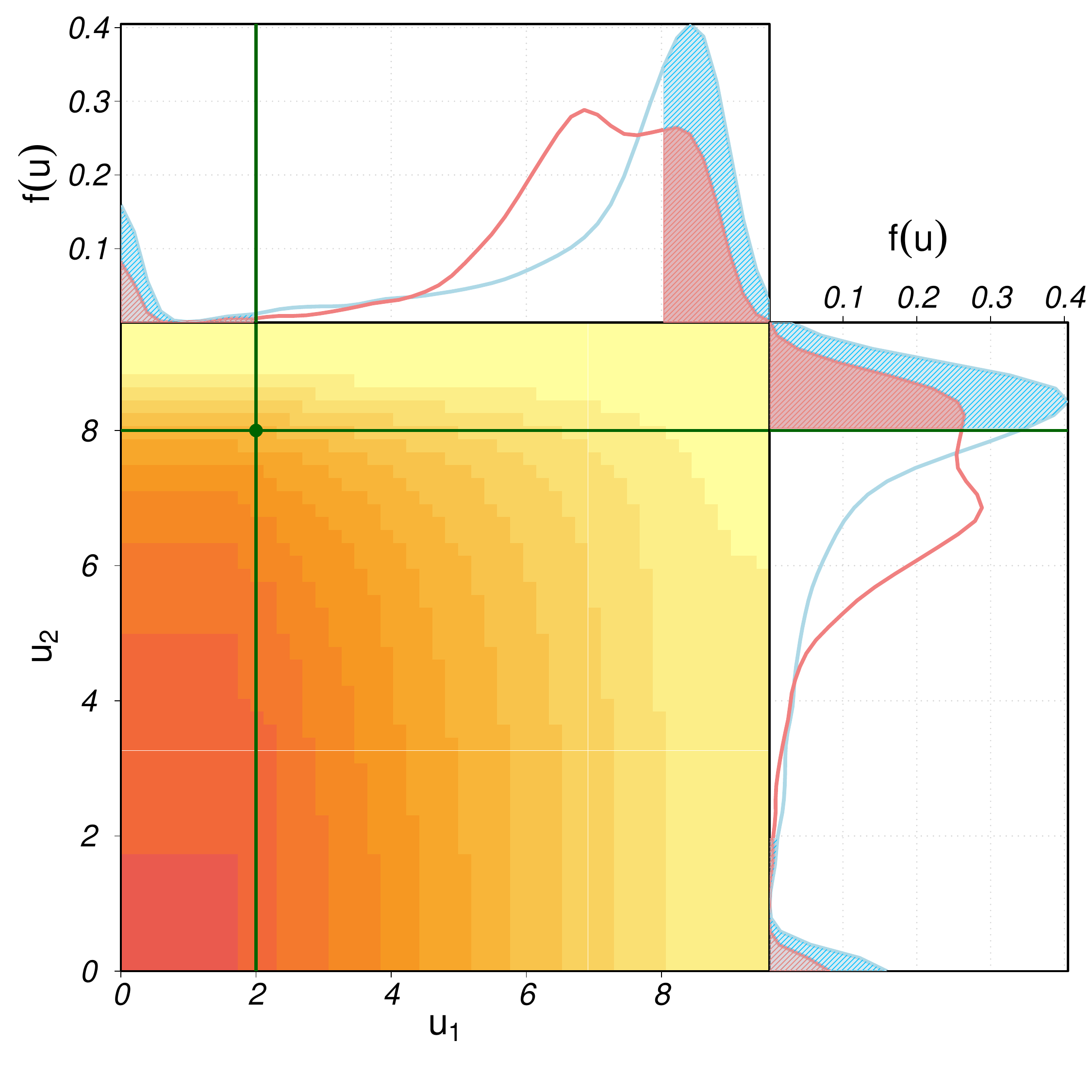}
         \caption{Subject 1: Age 47 years, EDSS = 1}
     \end{subfigure}
     \hfill
    \begin{subfigure}[b]{0.49\textwidth}
         \centering
         \includegraphics[width=\textwidth]{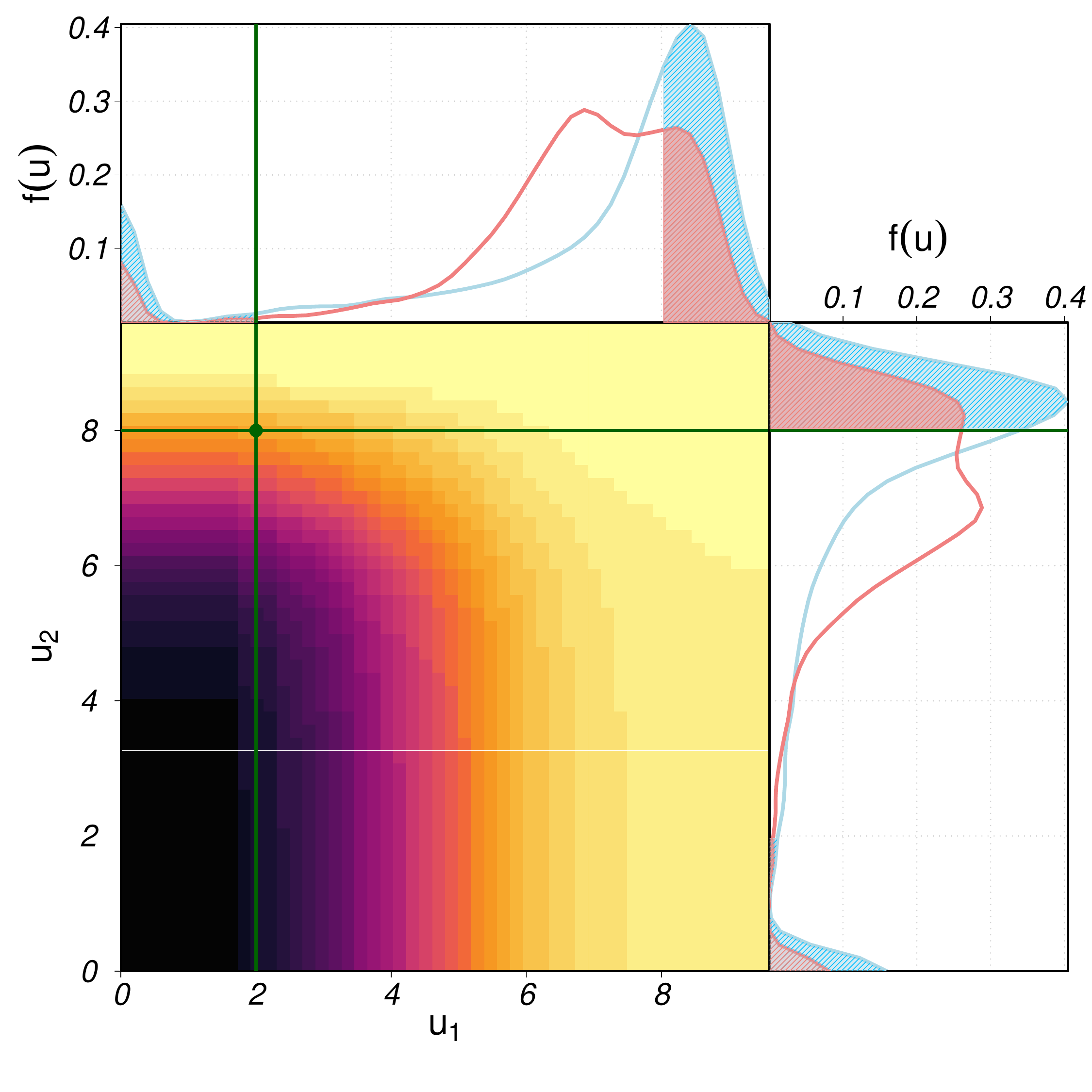}
         \caption{Subject 2: Age 62 years, EDSS  = 6}
     \end{subfigure}  
     \includegraphics[width=\textwidth]{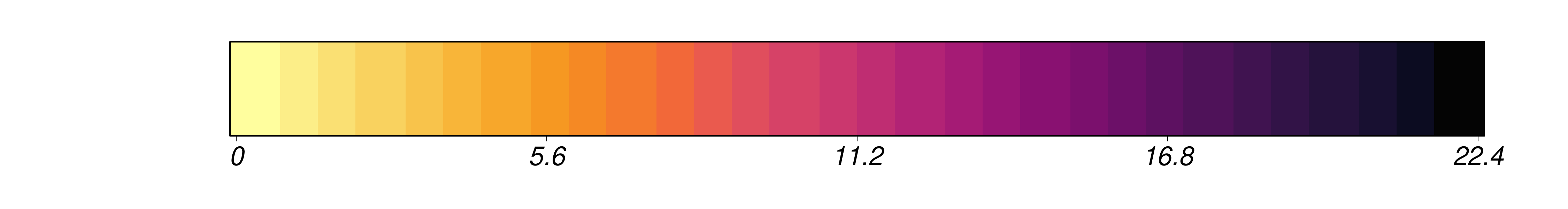}
     
     \caption{In panels (a) and (b), the top and right figures depict the density functions of the log-activity for Subjects 1 and 2, where light blue and light red represent Subjects 1 and 2, respectively. In each panel, the image plot illustrates the proposed 2-index function $h_{2, F}$. The green solid vertical line marks the sedentary activity level, while the green solid horizontal line represents the moderate-to-vigorous activity level. In the image plot, colors with higher identity indicate a higher values of $h_{2, F}$. \label{fig:odds2d}}
\end{figure}

Now, consider a function $h_{2, F}: [0, D]\times[0, D] \rightarrow \bbR_{\geq 0}$ such that 
\begin{equation}
    h_{2, F}(u_{1}, u_{2}) = \frac{S(u_{2})}{F(u_{1})} = \frac{\bbP\{X > u_{2}\}}{\bbP\{X \leq u_{1}\}}.
\end{equation}
Similar to the 1-index model, now we assume that the data $(y_{i}, h_{2, F_{i}}(\cdot))_{i=1}^{n}$ are random samples from $(y, h_{2,F})$. Then for a two-dimensional smoothed function $\beta_{2}(u_{1}, u_{2})$, the linear functional $\int_{0}^{D}\int_{0}^{D}\beta_{2}(u_{1}, u_{2})h_{2, F_{i}}(u_{1}, u_{2})du_{1}du_{2}$ describes the common effects of the entire trajectories $h_{2, F_{i}}$ on $y_{i}$. Figure \ref{fig:odds2d} demonstrates $h_{2, F_{i}}$ as a two-dimensional object for two subjects that the considered in Section \ref{sec:1index}. Note that, $h_{2, F}(u, u) = h_{1, F}(u)$, $h_{2, F}(u, D) = 0$ for all $u > 0$ and $h_{2, F}(D, u) = S(u)$. 
\par
Modeling such a two-dimensional object is more difficult than what we have discussed in Section \ref{sec:1index}. The functional linear model can be stated as 
\begin{equation}
\label{eq:model-2d}
    y_{i} \sim EF(\mu_{i}, \sigma^{2}) \qquad g(\mu_{i}) = \alpha_{2} + \int_{0}^{D}\int_{0}^{D}\beta_{2}(u_{1}, u_{2})h_{2, F_{i}}(u_{1}, u_{2})du_{1}du_{2}, 
\end{equation}
where $EF(\mu_{i}, \sigma^{2})$ is the exponential family with mean $
\mu_{i}$ and finite dispersion parameter $\sigma^{2} > 0$. In the above model, $\alpha_{2}$ is the scalar intercept. Now, we express the $g_{2}$ as 
\begin{align*}
    g_{2}(\mu_{i}) &= \alpha_{2} + \int_{0}^{D}\int_{0}^{D}\beta_{2}(u_{1}, u_{2})h_{2, F_{i}}(u_{1}, u_{2})du_{1}du_{2}\\
    &= \alpha_{2} +  \int_{0}^{D}\int_{0}^{D}
    \sum_{k_{1}=1}^{\kappa_{1}} 
    \sum_{k_{2}=1}^{\kappa_{2}} 
    b_{k_{1}, k_{2}}^{(2)}B_{k_{1}}(u_{1})B_{k_{2}}(u_{2})h_{2, F_{i}}(u_{1}, u_{2})du_{1}du_{2}\\
    &= \alpha_{2} + \sum_{k_{1}=1}^{\kappa_{1}}\sum_{k_{2}=1}^{\kappa_{2}} b_{k_{1}, k_{2}}^{(2)}W_{i, k_{1}, k_{2}}^{(2)}\\
    &= \alpha_{2} + \left<\bB^{(2)}, \bW_{i}^{(2)}\right>_{F}
    \numberthis
\end{align*}
where $W_{i, k_{1}, k_{2}}^{(2)} = \int_{0}^{D}\int_{0}^{D}B_{k_{1}}(u_{1})B_{k_{2}}(u_{2})h_{2, F_{i}}(u_{1}, u_{2})du_{1}du_{2}$ for $k_{1} = 1, \cdots, \kappa_{1}, k_{2} = 1, \cdots, \kappa_{2}$ and $i=1, \cdots, n$. Thus, the objective is to estimate the matrix $\bB^{(2)} = (b_{k_{1}, k_{2}}^{(2)})_{k_{1}, k_{2}}$ and hence $\beta_{2}(\cdot, \cdot)$.

\par
The two simple examples in Section \ref{sec:1index} and \ref{sec:2index} motivate us to generalize the odd functions and use them to model the scalar outcomes $y_{i}$s.

\subsection{Generalized odds}
\label{sec:genodds}
Based on the discussion in Section \ref{sec:framework}, we can generalize two odd functions defined in Sections \ref{sec:1index} and \ref{sec:2index}. For $u_{k} \in [0, D]$ for $k = 1, 2, 3, 4$, we can define $A = [u_{1}, u_{2}]$ and $B = [u_{3}, u_{4}]$ with the feasible constants $u_{1} < u_{2}$ and $u_{3} < u_{4}$. Therefore, the subject-specific generalized odd can be defined as the ratio of the probabilities of two events $\{X_{i} \in [u_{1}, u_{2}]\}$ and $\{X_{i} \in [u_{3}, u_{4}]\}$.   
\par
\begin{figure}
    \centering
    \includegraphics[width=\linewidth]{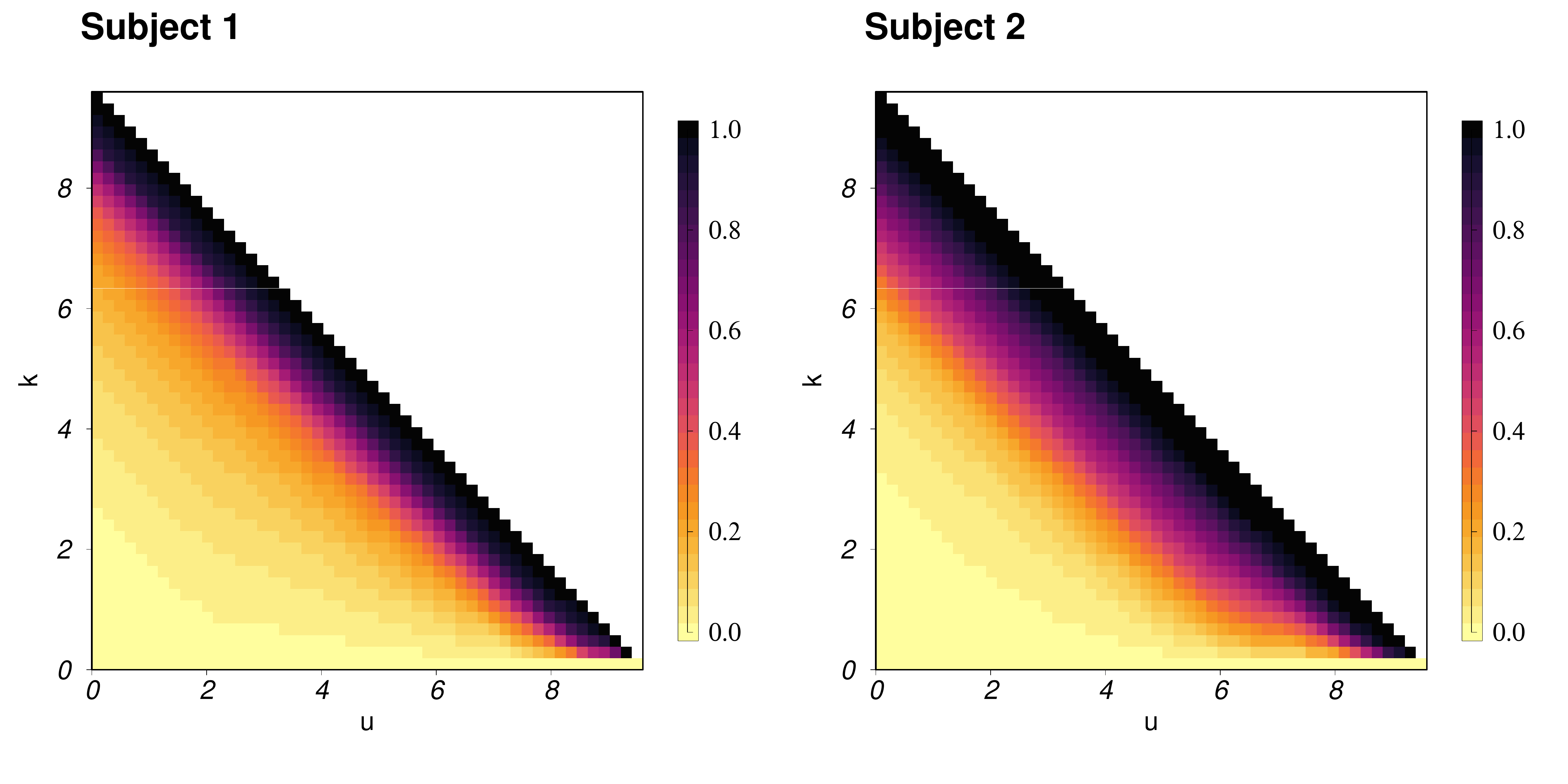}
    \caption{Residual life distribution for two subjects. Subject 1: 47 years old female with EDSS = 1 (light blue). Subject 2: 62 years old female with EDSS = 6 (light red).}
    \label{fig:rld}
\end{figure}

Let us introduce a subject-specific generalized odd (GO) (possibly smooth) function $h_{4, F}: [0, D]^{4} \rightarrow \bbR_{\geq 0}$ that produces different well-known probabilistic features for different values of its arguments such that 
\begin{equation}
\label{eq:h-general}
    h_{4, F}(u_{1}, u_{2}, u_{3}, u_{4}) = \left|\frac{F(u_{4}) - F(u_{3})}{F(u_{2}) - F(u_{1})}\right|,
\end{equation}
where $F$ is the distribution of the underlying random variable.  
\begin{remark}
    For different values/constants on $u_{k}$s, the generalized odd defined on Equation \eqref{eq:h-general} provides some well-known distributional representations. For example, $h_{4, F}(0, D, k, k-1) = p(k)$, the probability mass function; $h_{4, F}(k, D, k-1, k) = p(k)/S(k) = \lambda(k)$, hazard function at $k$ and $h_{4, F}(t, D, t, t+u) = \frac{F(t+u) - F(t)}{1-F(t)}$,
    residual life distribution for $x > 0$, where $h_{4, F}$ is the conditional distribution function of $X - t \leq u$ given that $X > t$. Figure \ref{fig:rld} demonstrates the residual life distribution for the same two subjects obtained through $h_{4, F}$. 
\end{remark}

Now consider the scalar-on-functional regression model where subject-specific mean is $\mu_{i} = \E\{Y_{i}|X_{i1}, X_{i2}, \cdots, X_{in_{i}}\}$. Therefore, for a known link function $g$,
\begin{equation}
\label{eq:model-4d}
     g(\mu_{i}) = \alpha + \int_{\sU}\beta(\bu)
     h_{F_{i}}(\bu)d\bu.
\end{equation}
In this article, we follow a smoothing spline technique that produces a good approximation of the unknown functional-coefficient $\beta$ in terms of the product of univariate basis functions expansion,
\begin{equation}
\label{eq:basis}
\beta(u_{1}, u_{2}, u_{3}, u_{4}) = 
\sum_{k_{1}=1}^{\kappa_{1}}\sum_{k_{2}=1}^{\kappa_{2}}
\sum_{k_{3}=1}^{\kappa_{3}}\sum_{k_{4}=1}^{\kappa_{4}}
b_{k_{1}, k_{2}, k_{3}, k_{4}}B_{k_{1}}(u_{1})B_{k_{2}}(u_{2})B_{k_{3}}(u_{3})B_{k_{4}}(u_{4}),
\end{equation}
where $\{B_{d}(u_{d})\}_{k = 1}^{\kappa_{d}}$ is the set of known basis functions over $u$ for $d = 1, 2, 3, 4$. We use a B-spline basis function that has desirable local properties. 
Now, based on the basis expansion of $\beta$, we can rewrite Equation \eqref{eq:basis} as follows.
\begin{align*}
\label{eq:g}
    g(\mu_{i}) = \alpha + 
    \sum_{k_{1}=1}^{\kappa_{1}}\sum_{k_{2}=1}^{\kappa_{2}}
    \sum_{k_{3}=1}^{\kappa_{3}}\sum_{k_{4}=1}^{\kappa_{4}}
    b_{k_{1}, k_{2}, k_{3}, k_{4}}W_{i, k_{1}, k_{2}, k_{3}, k_{4}},
    \numberthis
\end{align*}
where 
\begin{equation}
\label{eq:W}
    W_{i, k_{1}, k_{2}, k_{3}, k_{4}} = \int_{\sU}
    B_{k_{1}}(u_{1})B_{k_{2}}(u_{2})B_{k_{3}}(u_{3})B_{k_{4}}(u_{4})
    h_{F_{i}}(\bu)d\bu.
\end{equation}
Therefore, combining Equations \eqref{eq:g} and \eqref{eq:W}, we can rewrite the model \eqref{eq:g} as follows. 
\begin{equation}
\label{eq:g-gen}
    g(\mu_{i}) = \alpha + \left<\sfB, \sfW_{i} \right>
\end{equation}
In the above equation, $\left< \cdot, \cdot \right>$ indicates the inner product between two tensors with same dimensions; this can also be presented as $\left< \sfB , \sfW_{i} \right> = \sum\limits_{k_{1}=1}^{\kappa_{1}}\sum\limits_{k_{2}=1}^{\kappa_{2}}
\sum\limits_{k_{3}=1}^{\kappa_{3}}\sum\limits_{k_{4}=1}^{\kappa_{4}}
b_{k_{1}, k_{2}, k_{3}, k_{4}}W_{i, k_{1}, k_{2}, k_{3}, k_{4}}$, 
where $\sfB, \sfW_{i}$ are the tensors of ordered $\kappa_{1}\times\kappa_{2}\times\kappa_{3}\times\kappa_{4}$ for $i = 1, \cdots, n$. 

\section{Estimation methods}
\label{sec:estimation}

Based on the models described in Equation \eqref{eq:model-1d}, \eqref{eq:model-2d} and \eqref{eq:model-4d} in Section \ref{sec:model}, the object is to estimate the corresponding regression coefficients. Instead of estimating the infinite-dimensional regression coefficient function directly, we use the basis expansion of the underlying regression coefficient function with a known basis function. Therefore, the problem is to estimate the unknown basis coefficients based on a generalized linear model. In this section, we re-frame and introduce the estimation of the basis coefficient $\sfB$ in the model \eqref{eq:g-gen} to obtain the regression coefficient $\beta(\cdot)$ in the scalar-on-generalized odds described in Section \ref{sec:genodds}. 
\par
First, it is easy to observe that Equation \eqref{eq:g-gen} can be re-written as $g(\mu_{i}) = \alpha + \left< \sfB,\sfW_{i}\right> = \alpha + \bW_{i}^{\tp}\btheta$, where $\bW_{i} = \vect(\sfW_{i})$, a $\prod\limits_{d=1}^{4}\kappa_{d}$-dimensional stacked vector of the elements $\sfW_{i, k_{1}, k_{2}, k_{3}, k_{4}} = \int B_{k_{1}}(u_{1})B_{k_{2}}(u_{2})B_{k_{3}}(u_{3})B_{k_{4}}(u_{4})h_{F_{i}}(\bu)d\bu$, and $\btheta ( = \vect(\sfB))$ is the corresponding vector of unknown basis coefficients. Thus, the model \eqref{eq:g-gen} can be considered as a generalized linear model with subject-specific covariates $\bW_{i}$ and the objective is to estimate the unknown $\btheta$. Under this set-up, we use the penalized log-likelihood criterion with a specific penalty on the basis coefficients $\btheta$, which simultaneously identifies with distributional representations that are more informative. 
\begin{equation}
\label{eq:obj}
    \sR(\alpha, \beta(\cdot)) = -2\log\sL(\alpha, \btheta; Y_{i}, \bW_{i}) + \sum_{j=1}^{\kappa}\sP(|\theta_{j}|; \lambda)
\end{equation}
where $\sP$ is the scalar penalty function, $\lambda$ is the generic tuning parameter and $\kappa = \prod\limits_{d=1}^{4}\kappa_{d}$. This helps us to reduce the number of parameters in the model described in Equation \eqref{eq:g-gen} and allow a sparse representation of the scalar-on-generalized odds regression framework. 
\par
In general, penalty functions can be chosen from the set of popular variable selection methods described below. 
\begin{enumerate}[label=(\alph*)]
    \item\label{item:lasso}  Least absolute shrinkage and selection operator (popularly known as lasso) \citep{tibshirani1996regression} with $\sP(|\theta|; \lambda) = \lambda |\theta|$. It is important to note that the lasso estimator introduces a non-negligible bias in estimating the nonzero coefficients \citep{fan2001variable}, and due to this incurrence of bias, the lasso does not have the oracle property \citep{zou2006adaptive}.
    \item\label{item:elnet} The elastic net \citep{zou2005regularization} with with $\sP(\theta; \lambda, \alpha) = \lambda\left[ (1-\alpha)|\theta|^{2}/2 + \alpha |\theta|\right]$. In many applications, the covariates are often highly correlated; in such a situation, the ridge penalty poorly determines the regression coefficient, whereas the lasso penalty does not discriminate among correlated variables when performing variable selection. The elastic net penalty addresses this issue by encouraging averaging of highly correlated variables through its first term while promoting sparsity in the coefficients of those averaged variables through its second term. 
    \item\label{item:scad} Smoothly clipped absolute deviation (SCAD) \citep{fan2001variable} with $\sP(|\theta|; \lambda, a) = \lambda|\theta|\textbf{1}\left\{|\theta| \leq \lambda\right\} - \left(\frac{|\theta|^{2} - 2a\lambda|\theta| + \lambda^{2}}{2(a-1)}\right)\textbf{1}\left\{\lambda < |\theta| \leq a\lambda\right\} + \frac{(a+1)\lambda^{2}}{2}\textbf{1}\left\{|\theta| > a\lambda\right\}$ for $a \geq 2$. The SCAD penalty is continuously differentiable on $(-\infty, 0) \cup (0, \infty)$ and singular at zero with its derivatives zero outside the range $[-a\lambda_{n}, a\lambda_{n}]$. Therefore, the small coefficients are set to zero, and a few other coefficients are shrunk to zero while retaining the large ones as they are. As a result, the penalty smooths the estimate and ensures its unbiasedness.
    \item\label{item:mcp} Minimax Concave Penalty (MCP) \citep{Zhang2010} with $\sP(|\theta|; \lambda, \gamma) = \{ \lambda|\theta| - \theta^{2}/(2\gamma)\}\textbf{1}\{|\theta| \leq \gamma\lambda\} + \{\gamma\lambda^{2}/2\}\textbf{1}\{|\theta| > \gamma\lambda\}$, where $\lambda > 0$ is the tuning parameter that controls the sparsity and $\gamma > 1$ is a concavity parameter controlling how the nonconvex penalty is. . 
\end{enumerate}
\par
\par

\section{Data analysis}
\label{sec:data-analysis}
In this section, we thoroughly examine the HEAL-MS analyses to showcase the proposed approach's practicality based on the log-transformed activity counts. Within this segment, we maintain a constant number of grid points in each direction at 50. Furthermore, we set $D$ to be 9.6, resulting in a range of $[0, 9.6]$ along each direction. The functional objects are constructed using the B-spline basis function, which is a common practice for non-periodic functions. This is a piecewise polynomial function with continuity and smoothness constraints at the points where the pieces meet (called knots). B-spline basis functions have the advantages of very fast computation and great flexibility. We choose the degree of the polynomial pieces $q = 3$ and the number of interior points $L_{d} = 8$, therefore, the number of basis functions $\kappa_{d} = L_{d} + q + 1 = 12$ for $d = 1, 2, 3, 4$.
\par
We start with the basic models with the average activity counts as covariates in the simple linear regression. Along with the scalar summary, such as average activity counts, use a scalar-on-function regression model with subject-specific survival function as a functional covariate. These basic key findings are also available in \citet{niyogi2024hazard}. Subsequently, the subject-specific 1-index ($h_{1,F}$), 2-index ($h_{2,F}$), and 4-index ($h_{4,F}$) quantities are utilized as functional objects, as defined in Section~\ref{sec:model}. 
To ensure the robustness of our assessment, we apply a rigorous 5-fold cross-validation procedure with 100 random replications. The average cross-validated $R^{2}$, along with a 95\% confidence interval, is reported in the Table \ref{tab:cvR2} to provide a comprehensive evaluation of the model's performance.
\par

\begin{table}[h]
    \centering
    \caption{Cross-validated $R^{2}$s are also shown for predicting EDSS scores.}
    \scalebox{0.9}{
    \begin{tabular}{c|r|ccc}
        \toprule
        & & \multicolumn{3}{c}{Cross-validated $R^{2}$}\\ 
        \midrule
        & $\beta_{\text{AC[8am-8pm]}}\overline{\text{LAC}}_{i\text{[8am-8pm]}}$
        & \multicolumn{3}{c}{0.095 (0.082, 0.117)}\\ 
        \hdashline
        &$\int \beta_{S}(u)S_{i}(u)du$ && 0.115 (0.090, 0.142) &\\
        \midrule
        && 1-index & 2-index & 4 index\\
        \midrule
        \multirow{7}{*}{\rotatebox[origin=c]{90}{\quad\qquad Penalty}} 
        & lasso    & 0.095 (0.066, 0.125) & 0.151 (0.116, 0.182) & 0.208 (0.174, 0.234)\\
        & elnet    & 0.098 (0.075, 0.126) & 0.146 (0.103, 0.178) & 0.209 (0.170, 0.236)\\
        & scad     & 0.090 (0.062, 0.119) & 0.128 (0.081, 0.172) & 0.208 (0.168, 0.237)\\
        & mcp      & 0.089 (0.063, 0.118) & 0.131 (0.094, 0.168) & 0.208 (0.172, 0.238)\\
        \bottomrule
    \end{tabular}
    }
    \label{tab:cvR2}
\end{table}

\section{Discussion}

We introduced a generalized odds framework and demonstrated its utility in scalar-on-distribution regression. Methodologically, the proposed scalar-on-odds model represents an advance over existing distributional regression approaches by using subject-specific odds functions as covariates, thereby directly targeting distributional tail behavior. Most prior methods rely on single-index representations such as densities or quantile functions, which summarize overall distributional shape but do not explicitly contrast extreme versus typical values. In contrast, odds functions compare probabilities of extreme observations against more typical ones, yielding tail-focused biomarkers. This distinction translated into substantial performance gains in our MS accelerometry application: the most flexible four-index odds model achieved a cross-validated $R^2$ of approximately 0.21, nearly doubling the performance of a single-index survival-based model.

The most general odds specification considered here is a multidimensional functional object. Incorporating such rich covariates necessitates additional modeling structure, which we address using spline-based representations and structured regularization. In addition, rather than relying on direct estimation of densities which can be unstable when extreme observations are rare, the GO formulation aggregates tail information into probability ratios that remain well behaved even under limited support.

In summary, while our motivating application focused on wearable physical activity and MS disability, the proposed framework is broadly applicable to other domains where contrasts between extreme and typical behavior are scientifically meaningful. An important direction for future work is to extend this framework to settings in which generalized odds functions serve as outcomes rather than predictors.



\newpage
 
\par

\bibliographystyle{unsrtnat}
\bibliography{references}  

\end{document}